  \documentstyle[preprint,eqsecnum,aps]{revtex}
\begin{document}
\tightenlines
\draft
\preprint{HEP/123-qed}
\title{Fine structure of alpha decay in odd nuclei}
\author{M. Mirea}
\address{ Institute of Physics and Nuclear Engineering, 
Tandem Lab., P.O. Box MG--6,
Bucarest, Romania}
\date{\today}
\maketitle
\begin{abstract}

Using an $\alpha$ decay  level scheme,  the fine
structure in odd  nuclei is explained by taking into account the
radial and rotational couplings between the unpaired valence
nucleon and the 
core of the decaying system. It is shown that the experimental behavior of the
$\alpha$ decay fine structure phenomenon is governed by the
dynamical characteristics of the system. 
\end{abstract}

$~~~~~~~~~~~~~~~$Keywords: alpha decay fine structure


\narrowtext


The $\alpha$ decay fine structure was discovered by Rosenblum \cite{r1} since 
1929 by measuring the ranges of the emitted particle in the air.
Usually, theoretical attempts to investigate this phenomenon 
are based on the calculation of 
 the overlaps between, on one hand, 
the ground--state wave function of the parent
and, on the other hand, the antisymmetric product 
between the wave functions
of the nascent fragments in different configurations after the
scission \cite{mang,mang2}. However, quantitatively this 
phenomenon was not explained
rigorously.

It was evidenced \cite{p1},
at least formally, 
that the significance of the emitted particle preformation 
 and that of its penetrability through the barrier 
calculated from the ground--state configuration of the parent
up to the  scission point
are equivalent. This equivalence gives a support in the
attempt to investigate 
the $\alpha$ decay process using
fission theories.

 Recently, a theory based on the Landau--Zener effect
was developed \cite{m1,m2,m3}  intending
to describe quantitatively the
cluster decay fine structure phenomenon. This new formalism
succeeds to reproduce the values of the fine structure hindrance factors 
for the $^{14}$C emission from $^{223}$Ra being in
 a better agreement than other microscopic theories \cite{mg}
with the experiment.
In this alternative description, the cluster decay fine
structure is caused by   the promotion of the 
valence unpaired nucleon
on some excited daughter levels during the disintegration
process. It was claimed that the same promotion effect can also
govern the fine structure in the case of $\alpha$ decay.
The first step in   such a treatment  
is the elaboration of a two--center realistic level diagram 
during the whole disintegration process, that means, starting 
from the parent  single--particle energy distribution 
 and following  the  
variations of the level energies
up to their asymptotic  configuration attributed to the 
separated daughter nucleus and the alpha particle.
In this picture, it is only intended to treat the
$\alpha$ cluster with a smooth potential in order to estimate
its influence upon the
levels of the daughter during the disintegration process, like
a polarization effect. 
The particle--core couplings are produced merely between levels
belonging to the nascent heavy nucleus, so that  the
influence of the potential attributed to the $\alpha$ particle
needs only to be simulated.
Of course, in this context, it is
not assumed that the oscillator well is an appropriate tool to 
describe an alpha nucleus. So,
to attain this purpose in a realistic manner, the superasymmetric
two--center shell model (STCSM) described in Ref. \cite{m4} and
 improved in Ref. \cite{m1} was modified
in order to reproduce 
the single--particle levels assumed to describe an 
$\alpha$ cluster in the final stage of the disintegration.
Consequently, the 
alpha oscillator stiffness $\hbar\omega_{2\alpha}$
was forced to vary
gradually from the usual STCSM  value 
$\hbar\omega_{2}=41A_{2}^{1/3}$ when the normalized
coordinate $R_{n}$=0,
to the value extracted from
Ref. \cite{l1} 
$\hbar\omega_{\alpha}$=21.8 MeV
when the scission point is reached, that means $R_{n}=1$. 
The nuclear shape parametrization being defined by two
intersected spheres of different radii, $R_{1}$ for the daughter
and $R_{2}$ for the emitted fragment, the single generalized
coordinate remains the elongation $R$, denoting the distance
between the centers of the nuclei, or the normalized
coordinate $R_{n}=(R-R_{i})/(R_{f}-R_{i})$, where $R_{i}=R_{0}-R_{2}$,
$R_{f}=R_{1}+R_{2}$ and $R_{0}$ denotes the radius of the parent.
Subsequently, another three  modifications were
introduced in the present version of the STCSM in  
order to obtain good $\alpha$--level energies associated with the
value $\hbar\omega_{\alpha}$=21.8 MeV.
The first term characterizing the mass-asymmetry along
the $\rho$--axis of Eq. (43) of Ref. \cite{m4} 
(proportional to $\hbar\omega_{2}-\hbar\omega_{1}$) was multiplied by
the ratio $(\omega_{2\alpha}-\omega_{1})/(\omega_{2}-\omega_{1})$.
Another mass-asymmetry term  along the $z$--axis 
was diagonalized using the
eigenvalues of the asymmetric two--center potential:
\begin{equation}
\begin{array}{rl}
<\nu',n_{\rho}',m',s'|{m_{0}\over 2}(\omega_{2\alpha}^{2}-\omega_{2}^{2})
(z-z_{2})^{2}|\nu,n_{\rho},m,s>=&
\int_{0}^{\infty}{m_{0}\over 2}(\omega_{2\alpha}^{2}-\omega_{2}^{2})
(z-z_{2})^{2}Z_{\nu_{2}'}Z_{\nu_{2}}dz\delta_{s's}\delta_{n_{\rho}'n_{\rho}}
\delta_{m'm}\\
 =&
{C_{\nu_{2}'}C_{\nu_{2}}\over 2 \alpha_{2}}I_{\nu_{2}',2,\nu_{2},\zeta_{0}'}
(\hbar\omega_{2\alpha}^{2}/\omega_{2}-\hbar\omega_{2})\delta_{s's}
\delta_{n_{\rho}'n_{\rho}}
\delta_{m'm}.\\
\end{array}
\end{equation}
In this way,  energy shifts in single--particle levels associated
to the change from $\omega_{2}$ to $\omega_{2\alpha}$ are produced, 
so that a good two--center oscillator  with
$\hbar\omega_{1}=41A_{1}^{1/2}$ and $\hbar\omega_{2}=21.8$ MeV
is obtained after the scission.
Finally, the spin -- orbit and $L^{2}$ interaction matrix elements 
associated to the light fragment were 
multiplied by the ratio $\omega_{2\alpha}/\omega_{2}$ in order
to be coherent with the formal definitions of these interaction
terms which assume  proportionality to $\hbar\omega_{2}$.
The signification of the notations can be found in Ref. \cite{m4}.

The spin--orbit and $L^{2}$ coefficients, together with the depths of the
potentials for the $^{211}$Po and $^{207}$Pb used in this work are 
$\kappa=5.75\times 10^{-2}$, $\eta=0.43$ and $V_{c}$=55.80 MeV, respectively.
These values for the smooth spherical oscillator potentials 
were deduced from a fit of the $^{208}$Pb 
single--particle energies given in Ref. \cite{j1}: 
$E_{1i_{11/2}}=$-3.16 MeV,
 $E_{2g_{ 9/2}}=$-3.94 MeV,  $E_{3p_{ 1/2}}=$-7.37 MeV,  
$E_{2f_{ 5/2}}=$-7.94 MeV,  $E_{3p_{ 3/2}}=$-8.36 MeV  and
 $E_{1i_{13/2}}=$-9.00 MeV. 
The single--particle energies of  $^{4}$He \cite{mi1} are: 
-12 MeV, 1.38 MeV, 1.68 MeV, 5.34 MeV,
10.43 MeV and 11.86 MeV for the levels 1$s_{1/2}$, 1$p_{3/2}$,
1$p_{1/2}$, 2$s_{1/2}$, 1$d_{5/2}$ and 1$d_{3/2}$, respectively.
A good fit for the lower energies in the frame of the oscillator model
with $\hbar\omega_{\alpha}$=21.8 MeV
was found to be: $\kappa$=4.51$\times 10^{-3}$, $\eta=$0 and $V_{c}$=52.91 MeV.

The STCSM level scheme is plotted in Fig. \ref{fig1}.
In the following, the condition of consistency is also achieved, 
the same shape
for the microscopical and phenomenological models involved.
The parent and the daughter
do not have pronounced deformations, so that their 
ground state nuclear shapes 
can be approximated  with spheres. 
The ground state configuration of $^{211}$Po is
$(\pi(h_{9/2})^{2}\nu(g_{9/2})^{1})_{9/2^{+}}$ \cite{ja1}.
For clarity, the levels of the parent will be labeled
with the superscript P, while the superscripts D and $\alpha$
will be used for the daughter and the $\alpha$ nuclei, respectively.
Up to now, the model evidences the variation of the levels
for a modification of the nuclear shape, and indicates
the origin of the nucleons belonging to the $\alpha$--particle.
As an interesting feature, the STCSM predicts that
the linked level 1$s_{1/2}^{\alpha}$ of
the $\alpha$ particle  emerges from the  orbital
1$g_{9/2}^{\rm P}$ of the spherical $^{211}$Po, which is 
deeply located in the parent potential
well.  The present formalism intends to explain the
fine structure by considering single--particle transitions
due to the radial and the rotational couplings. 
The levels with the same good quantum numbers
associated to some symmetries of the system cannot in 
general intersect, but exhibit quasi--crossings, or
pseudo--crossings, or avoided level crossings. The system
is characterized by an axial symmetry, therefore the good quantum numbers
are the projections of the nucleon spin $\Omega$. The radial coupling
causes transitions of the unpaired nucleon near the avoided 
level crossings. True crossings can also be  obtained
between levels characterized by different quantum numbers.
 Generally, the rotational coupling has a maximum
strength in the vicinity of the true crossings. Transitions
due to both couplings are taken into account in order to
explain the excitations of the unpaired nucleon. It can be
considered that the last unpaired neutron, initially located
in the orbital 2$g_{9/2}^{\rm P}$ has the same chance to choose
one of the levels with the projection $\Omega$ included between
1/2 and 9/2 when the nucleus starts to deform, therefore the occupation
probabilities are the same. Comparing the diagrams (a)--(d) of
Fig. \ref{fig2}, it is clear that only the level emerging
from $2g_{9/2}^{\rm P}$ with the spin projection $\Omega$=1/2 finally reaches
adiabatically   the 3$p_{1/2}^{\rm D}$ 
ground--state of the daughter,
which means,  a disintegration performed without
excitations. Due to the rotational coupling, a nucleon initially
in the state 2$g_{9/2}^{\rm P}$ $\Omega$=3/2 can jump in 
the state $\Omega$=1/2 during the disintegration, contributing
in a smaller measure to obtain finally the daughter ground--state.
Even by taking into account the rotational coupling, the other levels 
with $\Omega>$3/2 emerging from $2g_{9/2}^{\rm P}$
have a negligible contribution in the ground--state channel.
Moreover, if the Coriolis coupling is not taken into account
(using only the radial coupling), the
levels with $\Omega\ne$1/2 emerging from 2$g_{9/2}^{\rm P}$ attain
finally higher daughter levels with at least 3 MeV excess in energy
(for example the orbital 2$g_{9/2}^{\rm D}$ of the daughter). 
As the
penetrability decreases dramatically with the height of the
barrier, it becomes clear that the processes characterized by
a nucleon emerging from 2$g_{9/2}^{\rm P}$
$\Omega>$3/2 in the initial moment of the disintegration
are unlikely.
This discussion allows us to
fix the initial conditions for our process:
initially, the valence nucleon of the decaying system
can be considered in the state 2$g_{9/2}^{\rm P}$
with $\Omega$=1/2 and 3/2.  
To show how it is possible that the unpaired
nucleon arrives on allowed excited states of the daughter
due only to the radial effect, an
arrow in Fig. \ref{fig2} (a) indicates some avoided level crossings
between the levels which reach adiabatically the 3$p_{1/2}^{\rm D}$,
2$f_{5/2}^{\rm D}$ and 3$p_{3/2}^{\rm D}$ daughter orbitals. The transition
probabilities are strongly enhanced in the avoided level crossing regions
in accordance with the Landau--Zener effect. The transition 
probability between two adiabatic single--particle states
strongly depends  on the velocity of passage through the avoided
crossing regions $v_{\rm tun}$, and, implicitly on 
the tunneling time of the barrier.
When the velocity increases, the transition probability is
enhanced and the nucleon follows with a larger probability the
so called diabatic state.
Using the above description, it is intended to reproduce
the fine structure \cite{ja1} exhibited by the $\alpha$ decay        
of $^{211}$Po: 98.9 \% transitions to
the  1/2$^{-}$ (level 3$p_{1/2}^{\rm D}$)
 ground state
of the daughter, 0.55\% transitions   
to the  5/2$^{-}$ (level 2f$_{5/2}^{\rm D}$) 
first single--particle excited state
and 0.54\% transitions
to the 3/2$^{-}$ (level 3p$_{3/2}^{\rm D}$) 
second single--particle excited state.
 
As briefly mentioned, both radial and rotational couplings caused by the
relative motion between the nascent fragments \cite{t1,t2}
are taken into account in order to calculate the occupation
probabilities  of several levels
of the daughter by the unpaired nucleon. 
For simplicity, the effect due to the radial coupling
is considered to be well reproduced by the Landau--Zener
promotion mechanism \cite{c1,www,pp} in the avoided crossing regions.
Therefore, the fine structure of the process is strongly related
to the dynamic characteristics of the system.
The rotational or Coriolis coupling causes transitions
between two levels for which the value of $\Omega$
differs by one unit and it is proportional to the angular
momentum operator of the single--particle $j_{\pm}$ matrix
element. The STCSM provides the ingredients for calculating
the single--particle transitions probabilities due
to the Landau--Zener effect and to the Coriolis couplings:
the interaction energies between the diabatic states 
$\epsilon_{ij}^{\Omega_{k}}$ in the avoided crossing
regions, the diabatic level energies $\epsilon_{i}^{\Omega_{k}}$
(using spline interpolations) 
and the wave functions required to compute
$\langle \Omega_{k} |j_{\pm}| \Omega_{k} \mp 1\rangle$ as described
in Refs. \cite{t1,t2}. The behavior of these ingredients
are plotted in Fig. \ref{fig3}. The relative velocity between
the
nascent fragments can also be calculated quantum mechanically
using a method similar to that of the variation of
constants \cite{s1}, but in the following it will be considered
as a fit parameter as in Refs. \cite{m1,m2,m3}. To obtain the
final occupation probabilities of the daughter levels by the
unpaired nucleon, a system of differential coupled
equations must be solved:
\begin{equation}
\begin{array}{ll}
\dot{c}_{i}^{\Omega_{k}}=&{1\over i\hbar}\sum_{j\ne i}
     \epsilon_{ij}^{\Omega_{k}}
     \exp\left(i\alpha_{ij}^{\Omega_{k}\Omega_{k}}\right)c_{j}^{\Omega_{k}}+\\
   & {1\over i\hbar}\sum_{l}{\hbar^{2}\over 2B_{R}R^{2}}
     \sqrt{I(I+1)-\Omega_{k}(\Omega_{k}+1)}
    \mid \langle i,\Omega_{k}|j_{+}|l,\Omega_{k}-1\rangle \mid
     \exp\left(i\alpha_{il}^{\Omega_{k}\Omega_{k}-1}\right)
    c_{k}^{\Omega_{k}-1}+\\
   & {1\over i\hbar}\sum_{l}{\hbar^{2}\over 2B_{R}R^{2}}
     \sqrt{I(I+1)-\Omega_{k}(\Omega_{k}+1)}
    \mid \langle i,\Omega_{k}|j_{-}|l,\Omega_{k}+1\rangle \mid
     \exp\left(i\alpha_{il}^{\Omega_{k}\Omega_{k}+1}\right)c_{k}^{\Omega_{k}+1}
\end{array}
\label{ecu}
\end{equation}
with $\alpha_{ij}^{\Omega_{l}\Omega_{m}}=\int_{0}^{t}
(\epsilon_{i}^{\Omega_{l}}-\epsilon_{j}^{\Omega_{m}})dt/\hbar$, $B_{R}$ is
the effective mass along the elongation $R$ which was
taken approximatively equal to the reduced mass of the system and
$I$=9/2 is the total spin of the system. 
The time dependence in the above  equation can be
removed by using the relations $\dot{c}_{i}^{\Omega_{k}}=
v_{\rm tun}\partial c_{i}^{\Omega_{k}}/\partial R$
and $R=v_{\rm tun}t$. The coefficients
$(c_{j}^{\Omega_{k}})^{2}$ give the occupation probabilities
of the diabatic levels $\{j,\Omega_{k}\}$. To solve this system,
following the above discussion  and inspecting the
Fig. \ref{fig2}, it was considered that
it is sufficient to choose the initial conditions so that
the levels with $\Omega$=1/2 and 3/2 emerging from 2$g_{9/2}^{\rm P}$
have the same initial occupation probabilities, which means that the
equality  
 $(c_{g9/2}^{1/2})^{2}+(c_{g9/2}^{3/2})^{2}=1$ is fulfilled. Also,
by solving the system (\ref{ecu}), it is satisfactory to take
into account the levels with $\Omega$=1/2 emerging from
$1i_{11/2}^{\rm P}$, 2$g_{9/2}^{\rm P}$, 3$p_{1/2}^{\rm P}$, 
2$f_{5/2}^{\rm P}$, 3$p_{3/2}^{\rm P}$,
those with   $\Omega$=3/2 emerging from
 2$g_{9/2}^{\rm P}$, 2$f_{5/2}^{\rm P}$, 3$p_{3/2}^{\rm P}$, 
and those  with   $\Omega$=5/2 emerging from
 2$g_{9/2}^{\rm P}$, 1$i_{13/2}^{\rm P}$.  All the avoided crossing
levels between the selected adiabatic states are taken into account.
Levels with $\Omega>$5/2 do not reach the
final channels we are interested in.

For each channel, the penetrability $P_{L_{im_{i}}}(Q_{i})$
of the barrier was obtained using the numerical superasymmetric
fission model \cite{poe}, the nuclear part being given
by the Yukawa--plus--exponential approximation. This penetrability
depends on the $Q_{i}$--value of the channel $i$
($i$ labels here the single--particle state of the daughter)
and of the relative motion orbital momentum $L_{im_{i}}$.
In the final channel $3p_{1/2}^{\rm D}$, due to the conservation laws,
$L_{im_{i}}$ has the value 5 ($m_{i}$=1), in the final channel
2$f_{5/2}^{\rm D}$, $L_{im_{i}}$ can be either 3, 5 or 7 ($m_{i}$=1,2,3)
and in the final channel
3$p_{3/2}^{\rm D}$, $L_{im_{i}}$ can be either 3 or 5 ($m_{i}$=1,2).
For a specific final single--particle state, for example 2$f_{5/2}^{\rm D}$,
it is not possible to discriminate between the possible values
of the relative motion  orbital momentum $L_{im_{i}}$=3,5 and 7
in order to compute only one barrier penetrability. In these circumstances,
by analogy to  the Mang's formulae of Refs. \cite{mang,mang2} 
concerning the radial motion and the associated angular momentum,  
we consider that
 the angular momentum $L_{im_{i}}$,
used in calculating the penetrabilities, has a probability to
be obtained in the final channel directly proportional
to  the square of the Clebsh--Gordon coefficient
$(jI\Omega -\Omega|L_{im_{i}}0)^{2}$. So that, the spectroscopic
amplitude in the channel $i$ associated to the spin $\Omega_{k}$
and the momentum $L_{im_{i}'}$ will be  
$p_{i}^{\Omega_{k}L_{im_{i}'}}=(c_{i}^{\Omega_{k}})^{2}
(j_{i}I\Omega_{k}-\Omega_{k}|L_{im_{i}'}0)^{2}/
\sum_{m_{i}}(j_{i}I\Omega_{k}-\Omega_{k}|L_{im_{i}}0)^{2}$
where the summation on $m_{i}$ is done on the allowed values
of $L_{im_{i}}$. The partial half--life $T_{i}^{\Omega_{k}}$ for the channel
$\{i,\Omega_{k}\}$ becomes proportional to the quantity:
\begin{equation}
T_{i}^{\Omega_{k}}\propto {1\over\sum_{m_{i}} p_{i}^{\Omega_{k}L_{im_{i}}}
P_{L_{im_{i}}}(Q_{i})},
\end{equation}
the proportionality factor being given by the barrier assault
frequency. The partial half--lives for the transitions
to the ground--state $T_{3p_{1/2}^{\rm D}}$,
to the first excited state $T_{2f_{5/2}^{\rm D}}$
and to the second excited state $T_{3p_{3/2}^{\rm D}}$ are:
\begin{equation}
\begin{array}{l}
{1\over T_{3p_{1/2}^{\rm D}}}={1\over T_{3p_{1/2}^{\rm D}}^{1/2}}\\
{1\over T_{2f_{5/2}^{\rm D}}}={1\over T_{2f_{5/2}^{\rm D}}^{1/2}}+
{1\over T_{2f_{5/2}^{\rm D}}^{3/2}}+{1\over T_{2f_{5/2}^{\rm D}}^{5/2}}\\
{1\over T_{3p_{3/2}^{\rm D}}}={1\over T_{3p_{3/2}^{\rm D}}^{1/2}}+
{1\over T_{3p_{3/2}^{\rm D}}^{3/2}}
\end{array}
\end{equation}
The barrier assault frequency being the same for
all the channels, the relative intensities 
$T_{3p_{1/2}^{\rm D}}/T_{2f_{5/2}^{\rm D}}$ and
$T_{3p_{1/2}^{\rm D}}/T_{3p_{3/2}^{\rm D}}$
for the fine
structure can be obtained. Several tunneling velocities,
considered here as a fit parameter, have been tried. For a tunneling velocity
of 9$\times 10^{6}$ fm/fs, the ratio between the intensity
for transitions
to the first excited state and  to the ground state was found
to be 0.0071 and the obtained ratio of the same parameter between
the second excited state and the ground state was 
0.0062. These results are in good agreement with the
experimental values presented before. Moreover, calculations
of Ref. \cite{s1} show that in the quantum time--dependent
approach, the tunneling velocity is of the order of 1$\times 10^{7}$ fm/fs.
These calculations suggest that the $\alpha$ decay fine structure
phenomenon can be explained quantitatively by describing the
decaying system with molecular models, and it can be stated
that the quantitative characteristics of this phenomenon are
ruled by dynamical effects. 
In an avoided crossing region, the two eigenfunctions
of the adiabatic levels exchange their characteristics.
If the relative distance $R$ change infinitely slow, the
unpaired nucleon will remain in the same adiabatic
single particle state after the passage through a quasi--crossing,
any other available single particle excited state being
unfavoured. For a large tunneling velocity, the unpaired nucleon
will follow the diabatic state after the passage through a
quasi--crossing, all the other states being unfavoured. For
a finite tunneling velocity, an intermediate situation arises,
some single particle states being favoured and other
single particle states being unfavoured.
The model propose here, as the usual picture that consists of
calculating the overlaps between the parent and
the channel wave functions, is also based on  the
existence of favoured and unfavoured transitions in odd nuclei.
The proposed  formalism offers a competitive description of the
alpha decay mechanism, by investigating for the first time 
the modality in which
the levels initially bunched in shells are reorganized during
the disintegration to realize the final energy configuration.

\newpage
\begin{figure}
\caption{Neutron level scheme for the $\alpha$ decay of $^{211}$Po
as function of the normalized elongation.
 The spectroscopic notations are used 
to describe the levels of the parent 
 nucleus in the left side of the figure  
while in the right side of the figure the
levels of the daughter and the alpha particle are labeled in the
first and second column, respectively.}
\label{fig1}
\end{figure}

\begin{figure}
\caption{Detailed part of the level scheme. 
The levels with $\Omega=1/2$ (a), $\Omega=3/2$ (b),
$\Omega=5/2$ (c),  $\Omega=7/2$ (d), $\Omega=9/2$ (e) and 
$\Omega=13/2$ (f) are plotted with thick lines.
The levels with $\Omega=11/2$ (e) and $\Omega=15/2$ (f)
are plotted with thick dotted lines.
}
\label{fig2}
\end{figure}

\begin{figure}
\caption{Differences between the adiabatic levels with spin
$\Omega$=1/2 are presented in the plots (a) to (f) suggesting
the possible avoided crossing regions, that means,
points of nearest approach between two adiabatic levels.
These regions are marked with arrows and the spectroscopic
notations are displayed on each plot. The adiabatic $E_{i}$
and the diabatic $\epsilon_{i}$ levels are presented in
the picture (g) with full and dotted lines respectively
only in the case $\Omega$=1/2. The interaction energies $\epsilon_{ij}$
between the diabatic levels of the plot (g) are presented in
(h). The levels with spin $\Omega$=1/2 emerging
from 2$f_{5/2}^{\rm P}$, 3$p_{3/2}^{\rm P}$ and 
1$i_{13/2}^{\rm P}$ are presented as 
full lines while the level with spin 3/2 emerging from
1$i_{13/2}^{\rm P}$ is plotted as  dotted line in picture (i).
For the levels presented in  (i), the matrix
elements $\langle 3/2|j_{+}|1/2\rangle=\langle 1/2|j_{-}|3/2 \rangle$
are drawn in (j). Asymptotically, for $R_{n}$=0, the matrix element between
$\Omega$=1/2 and 3/2 of the levels belonging to 
the subshell 1$i_{13/2}^{\rm P}$ (full line) has the value
$\langle\Omega\pm 1|j_{\pm}|\Omega\rangle=
\hbar\sqrt{(j\mp\Omega)(j\pm\Omega+1)}$
while for $R_{n}\rightarrow\infty$ the same value is obtained within
the matrix element for the level $\Omega$=1/2 emerging from
$3p_{3/2}^{\rm P}$ and the level 1$i_{13/2}^{\rm P}$
$\Omega$=3/2 (dashed line). 
Otherwise, asymptotically the values
are zero (dotted line). A pronounced maximum of the matrix elements is
obtained when the levels with different spin projections 
intersect.}
\label{fig3}
\end{figure}


\begin{thebibliography}{99}

\bibitem{r1}    S. Rosenblum, C. R. Acad. Sci. Paris {\bf 188}, 1401 (1929).
\bibitem{mang}
H.J. Mang, Z. Phys. {\bf 148}, 528 (1957).
\bibitem{mang2}
H.J. Mang, Phys. Rev. {\bf 119}, 1069 (1960).
\bibitem{p1} D.N. Poenaru and W. Greiner, J. Phys. G {\bf 17}, S443 (1991).
\bibitem{m1}
M. Mirea, Phys. Rev. C {\bf 57}, 2484 (1998).
\bibitem{m2}
M. Mirea and F. Clapier, Europhy. Lett. {\bf 40}, 509 (1997).
\bibitem{m3}
M. Mirea, Eur. Phys. J. A {\bf 4}, 335 (1999).
\bibitem{mg}
M. Mirea and R.K. Gupta, in {\it Heavy Elements and Related New Phenomena},
Edited by. W. Greiner and R.K. Gupta  
(World Scientific, Singapore, 1999),
chap. 19.
\bibitem{m4}
M. Mirea, Phys. Rev. C {\bf 54}, 302 (1996).
\bibitem{l1}
R.D. Lawson, in {\it Theory of the Nuclear Shell Model},
(Clarendon Press, Oxford, 1980), p. 236.
\bibitem{j1}
C.H. Johnson, D.J. Horen and C. Mahaux,
Phys. Rev. C {\bf 36}, 2252 (1987).
\bibitem{mi1}
D.J. Millener and D. Kurath, Nucl. Phys. A {\bf 255}, 315 (1975).
\bibitem{ja1}
L.J. Jardine, Phys. Rev. C {\bf 11}, 1385 (1975).
\bibitem{t1} 
G. Terlecki, W. Scheid, H.J. Fink and W. Greiner,
Phys. Rev. C {\bf 18}, 265 (1978).
\bibitem{t2}
A. Thiel, W. Greiner, J.Y. Park and W. Scheid,
Phys. Rev. C {\bf 36}, 647 (1987).
\bibitem{c1}
M.H. Cha, J.Y. Park and W. Scheid, Phys. Rev. C {\bf 36},
2341 (1987).
\bibitem{www} W. Greiner, J. Y. Park and W. Scheid, in
{\it Nuclear Molecules} (World Scientific, Singapore, 1995), Chap. 11.
\bibitem{pp}
J.Y. Park, W. Greiner and W. Scheid, Phys. Rev. C {\bf 21}, 958 (1980).
\bibitem{s1} O. Serot, N. Carjan and D. Strottman, Nucl. Phys. A 
{\bf 569},
562 (1994).
\bibitem{poe} D.N. Poenaru and M. Ivascu, in {\it Particle
Emission from Nuclei}, edited by D.N. Poenaru and M.S. Ivascu
(CRC Press, Boca Raton, 1989), Vol. II, Chaps. 4 and 5.
\end{thebibliography}
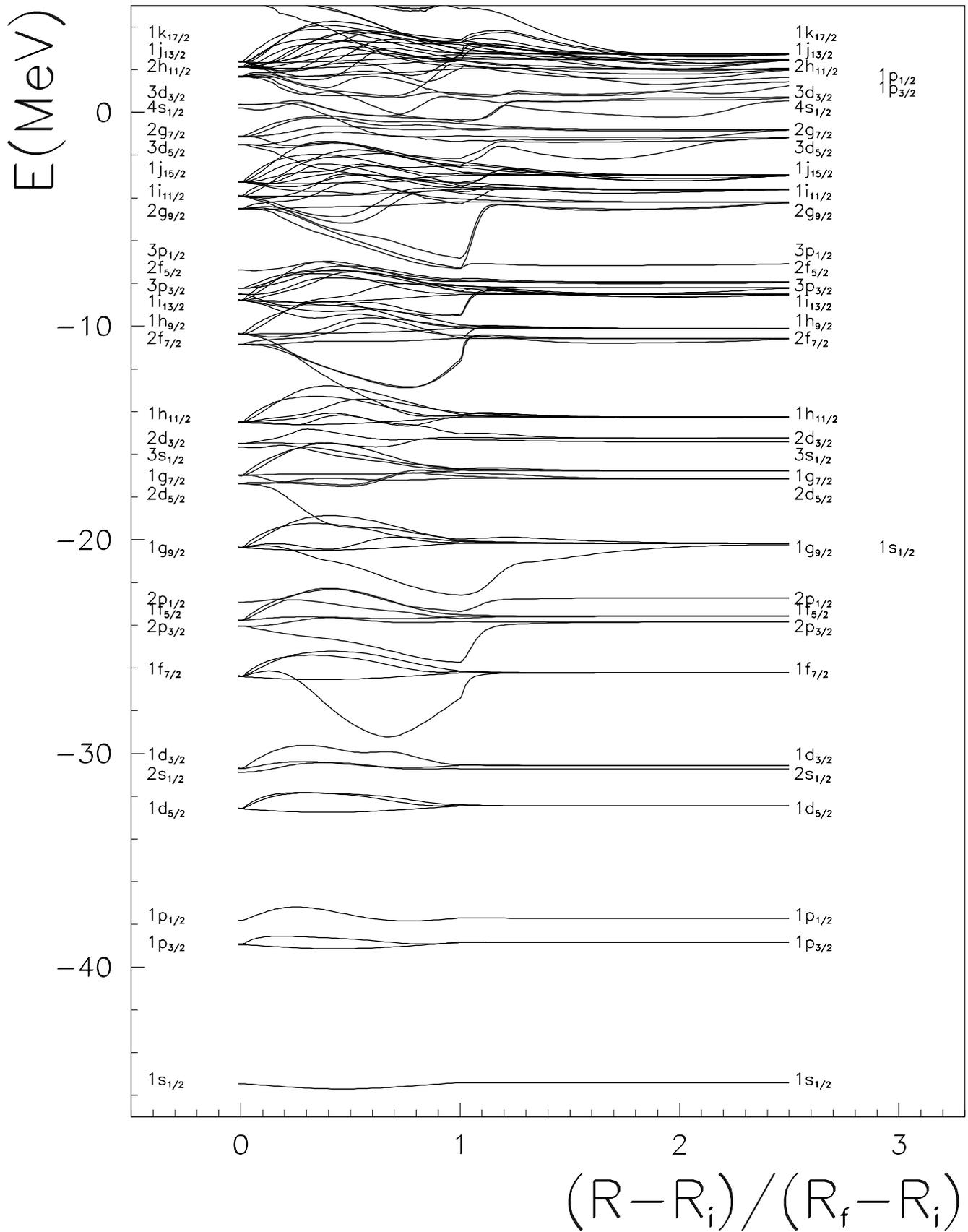
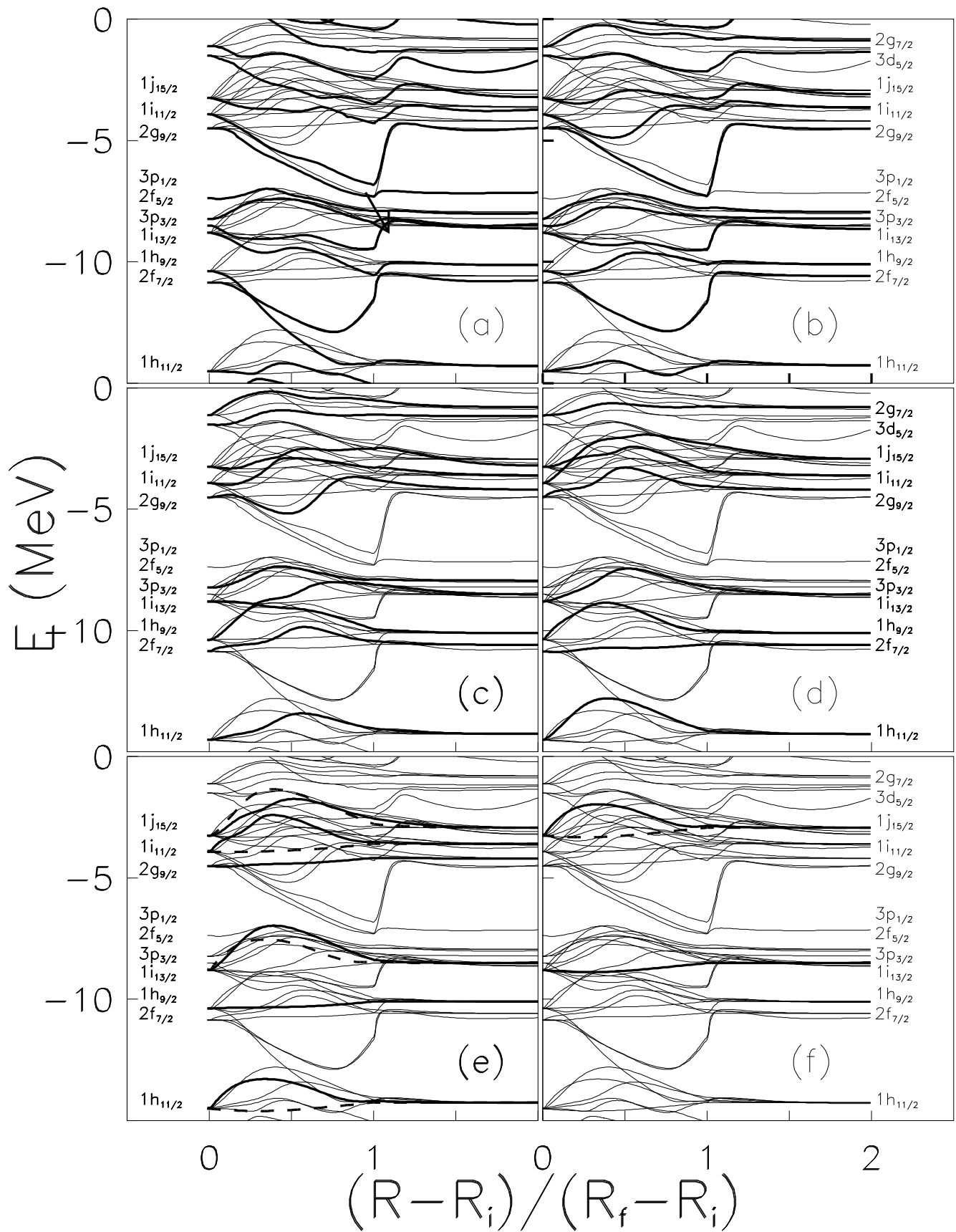
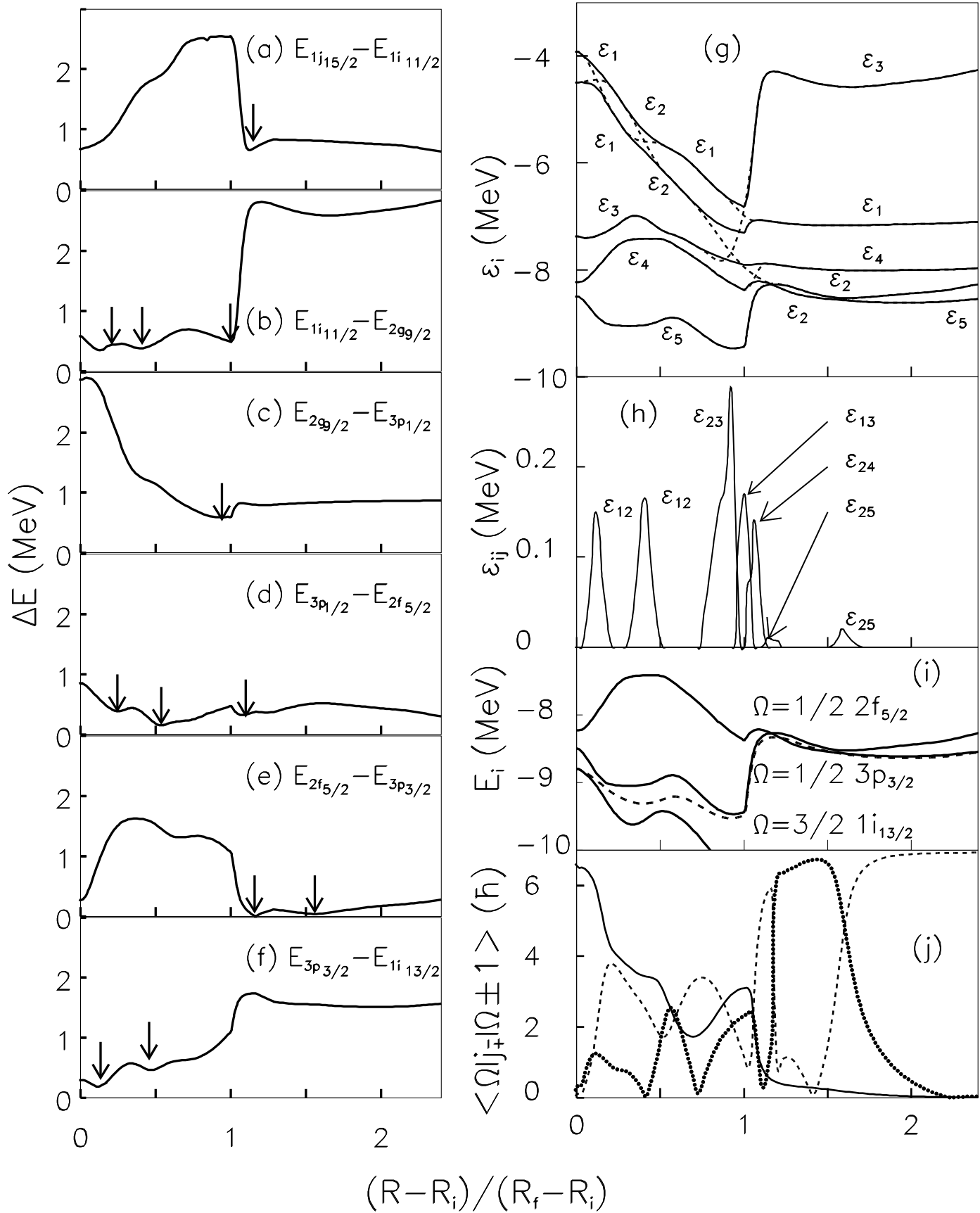
\end{document}